\documentstyle[12pt]{article}
\begin{document}

\thispagestyle{empty}
\renewcommand{\thefootnote}{\fnsymbol{footnote}}
\setcounter{footnote}{0}

\def\pxb{\left(\p \times \B - \B \times \p \right)}
\def\LAMBDA{\lambdabar}
\def\rk{r_k}
\def\beq{\begin{equation}}
\def\eeq{\end{equation}}
\def\bea{\begin{eqnarray}}
\def\eea{\end{eqnarray}}
\def\nn{\nonumber}
\def\ba{\begin{array}}
\def\ea{\end{array}}
\def\0{{\mbox{\boldmath $0$}}}
\def\one{1\hskip -1mm{\rm l}}
\def\A{{\mbox{\boldmath $A$}}}
\def\B{{\mbox{\boldmath $B$}}}
\def\El{{\mbox{\boldmath $E$}}}
\def\F{{\mbox{\boldmath $F$}}}
\def\S{{\mbox{\boldmath $S$}}}

\def\a{{\mbox{\boldmath $a$}}}
\def\p{{\mbox{\boldmath $p$}}}
\def\hatp{{\hat{\mbox{\boldmath $p$}}}}
\def\hatP{{\hat{\mbox{\boldmath $P$}}}}
\def\vpi{{\mbox{\boldmath $\pi$}}}
\def\hatvpi{\hat{\mbox{\boldmath $\pi$}}}
\def\r{{\mbox{\boldmath $r$}}}
\def\v{{\mbox{\boldmath $v$}}}
\def\w{{\mbox{\boldmath $w$}}}
\def\H{{\rm H}}
\def\hA{\hat{A}}
\def\hB{\hat{B}}
\def\i{{\rm i}}
\def\ih{\frac{\i}{\hbar}}
\def\ixh{\i \hbar}
\def\ddx{\frac{\partial}{\partial x}}
\def\ddy{\frac{\partial}{\partial y}}
\def\ddz{\frac{\partial}{\partial z}}
\def\ddt{\frac{\partial}{\partial t}}
\def\vsig{{\mbox{\boldmath $\sigma$}}}
\def\Al{{\mbox{\boldmath $\alpha$}}}
\def\ho{\hat{\cal H}_o}
\def\half{\frac{1}{2}}
\def\E{{\hat{\cal E}}}
\def\O{{\hat{\cal O}}}
\def\eps{\epsilon}
\def\g{\gamma}
\def\Vomeg{{\underline{\mbox{\boldmath $\Omega$}}}_s}
\def\hH{\hat{H}}
\def\Vsig{{\mbox{\boldmath $\Sigma$}}}
\def\Nab{{\mbox{\boldmath $\nabla$}}}
\def\curl{{\rm curl}}
\def\bh{\bar{H}}
\def\th{\tilde{H}}

\def\zone{z^{(1)}}
\def\ztwo{z^{(2)}}
\def\zi{z_{\rm in}}
\def\zo{z_{\rm out}}

\def\At{{\hat{A}(t)}}
\def\dAt{\frac{\partial {\hat{A}(t)} }{\partial t}}
\def\sone{{\hat{S}_1}}
\def\dsone{\frac{\partial {\hat{S}_1} }{\partial t}}
\def\dO{\frac{\partial {\hat{\cal O}}}{\partial t}}

\begin{center}
{\Large\bf Maxwell Optics: I. An exact matrix representation of the
Maxwell equations in a medium}

\bigskip

{\em Sameen Ahmed KHAN} \\

\bigskip

khan@fis.unam.mx ~~~ http://www.pd.infn.it/$\sim$khan/ \\
rohelakhan@yahoo.com ~~~ http://www.imsc.ernet.in/~jagan/khan-cv.html \\
Centro de Ciencias F\'{i}sicas,
Universidad Nacional Aut\'onoma de M\'exico, \\
Apartado Postal 48-3,
Cuernavaca 62251,
Morelos, \\
{\bf M\'EXICO} \\

\end{center}
\medskip
\medskip

\noindent
{\bf Abstract} \\
Matrix representations of the Maxwell equations are well-known.
However, all these representations lack an exactness or/and are given
in terms of a {\em pair} of matrix equations.  We present a
matrix representation of the Maxwell equation in presence of
sources in a medium with varying permittivity and permeability.
It is shown that such a representation necessarily requires
$8 \times 8$ matrices and an explicit representation for them
is presented.

\bigskip

\section{Introduction}
Matrix representations of the Maxwell equations are very
well-known~\cite{Moses}-\cite{Birula}.  However, all these
representations lack an exactness or/and are given in terms of a
{\em pair} of matrix equations.  Some of these representations are
in free space.  Such a representation is an approximation in a
medium with space- and time-dependent permittivity
$\epsilon (\r , t)$ and permeability $\mu (\r , t)$ respectively.
Even this approximation is often expressed through a pair of
equations using $3 \times 3$ matrices: one for the curl and one
for the divergence which occur in the Maxwell equations.  This
practice of writing the divergence condition separately is
completely avoidable by using $4 \times 4$ matrices~\cite{Moses}
for Maxwell equations in free-space.  A single equation using
$4 \times 4$ matrices is necessary and sufficient when
$\epsilon (\r , t)$ and $\mu (\r , t)$ are treated as `local'
constants~\cite{Moses,Birula}.

A treatment taking into account the variations of
$\epsilon (\r , t)$ and $\mu (\r , t)$ has been presented
in~\cite{Birula}.  This treatment uses the
Riemann-Silberstein vectors, $\F^{\pm} \left(\r , t \right)$
to reexpress the Maxwell equations as four equations: two
equations are for the curl and two are for the divergences and
there is mixing in
$\F^{+} \left(\r , t \right)$ and $\F^{-} \left(\r , t \right)$.
This mixing is very neatly expressed through the two derived
functions of $\epsilon (\r , t)$ and $\mu (\r , t)$.
These four equations are then expressed as a pair of matrix
equations using $6 \times 6$ matrices: again one for the curl and
one for the divergence.   Even though this treatment is exact it
involves a pair of matrix equations.

Here, we present a treatment which enables us to express the
Maxwell equations in a single matrix equation instead of a {\em pair}
of matrix equations.  Our approach is a logical continuation of the
treatment in~\cite{Birula}.  We use the linear combination of the
components of the  Riemann-Silberstein vectors,
$\F^{\pm} \left(\r , t \right)$ and the final matrix representation
is a single equation using $8 \times 8$ matrices.  This representation
contains all the four Maxwell equations in presence of sources taking
into account the spatial and temporal variations of the permittivity
$\epsilon (\r , t)$ and the permeability $\mu (\r , t)$.

In Section-I we shall summarize the treatment for a homogeneous medium
and introduce the required functions and notation.  In Section-II we
shall present the matrix representation in an inhomogeneous medium,
in presence of sources.

\section{Homogeneous Medium}
We shall start with the Maxwell
equations~\cite{Jackson, Ponofsky-Phillips} in an inhomogeneous medium
with sources,
\bea
\Nab \cdot {\mbox{\boldmath $D$}} \left(\r , t \right)
=
\rho\,, \nn \\
\Nab \times {\mbox{\boldmath $H$}} \left(\r , t \right)
- \frac{\partial }{\partial t}
{\mbox{\boldmath $D$}} \left(\r , t \right)
=
{\mbox{\boldmath $J$}}\,, \nn \\
\Nab \times {\mbox{\boldmath $E$}} \left(\r , t \right)
+
\frac{\partial }{\partial t}
{\mbox{\boldmath $B$}} \left(\r , t \right)
= 0\,, \nn \\
\Nab \cdot {\mbox{\boldmath $B$}} \left(\r , t \right)
= 0\,.
\label{Maxwell-1}
\eea
We assume the media to be linear, that is
${\mbox{\boldmath $D$}} = \epsilon \El$, and
${\mbox{\boldmath $B$}} = \mu {\mbox{\boldmath $H$}}$,
where $\epsilon$ is the {\bf permittivity of the medium} and
$\mu$ is the {\bf permeability of the medium}.  In general
$\epsilon = \epsilon (\r , t)$ and $\mu = \mu (\r , t)$.  In
this section we treat them as `local' constants in the various
derivations.  The magnitude of the velocity of light in the medium
is given by
$v (\r , t)  = \left|{\mbox{\boldmath $v$}} (\r , t)\right|
= {1}/{\sqrt{\epsilon (\r , t) \mu (\r , t)}}$.  In vacuum we
have, $\epsilon_0 = 8.85 \times 10^{- 12} {C^2}/{N. m^2}$ and
$\mu_0 = 4 \pi \times 10^{- 7} {N}/{A^2}$.

One possible way to obtain the required matrix representation is
to use the Riemann-Silberstein vector~\cite{Birula} given by
\bea
\F^{+} \left(\r , t \right)
& = &
\frac{1}{\sqrt{2}}
\left(
\sqrt{\epsilon (\r , t)} \El \left(\r , t \right)
+ \i \frac{1}{\sqrt{\mu (\r , t)}} \B \left(\r , t \right) \right) \nn \\
\F^{-} \left(\r , t \right)
& = &
\frac{1}{\sqrt{2}}
\left(
\sqrt{\epsilon (\r , t)} \El \left(\r , t \right)
- \i \frac{1}{\sqrt{\mu (\r , t)}} \B \left(\r , t \right) \right)\,.
\label{R-S-Conjugate}
\eea
For any homogeneous medium it is equivalent to use either
$\F^{+} \left(\r , t \right)$ or $\F^{-} \left(\r , t \right)$.  The
two differ by the sign before `$\i$' and are not the complex conjugate
of one another.  We have not assumed any form for $\El (\r , t)$ and
$\B (\r , t)$.  We will be needing both of them in an inhomogeneous
medium, to be considered in detail in Section-III.

If for a certain medium $\epsilon (\r , t)$ and $\mu (\r , t)$ are
constants (or can be treated as `local' constants under certain
approximations), then the vectors $\F^{\pm} \left(\r , t \right)$
satisfy
\bea
\i \frac{\partial }{\partial t} \F^{\pm} \left(\r , t \right)
& = &
\pm v \Nab \times \F^{\pm} \left(\r , t \right)
- \frac{1}{\sqrt{2 \epsilon}} (\i {\mbox{\boldmath $J$}}) \nn \\
\Nab \cdot \F^{\pm} \left(\r , t \right)
& = &
\frac{1}{\sqrt{2 \epsilon}} (\rho)\,.
\label{RS-F-Equations-Approximate}
\eea
Thus, by using the Riemann-Silberstein vector it has been possible
to reexpress the four Maxwell equations (for a medium with constant
$\epsilon$ and $\mu$) as two equations.  The first one contains the
the two Maxwell equations with curl and the second one contains the
two Maxwell with divergences.  The first of the two equations
in~(\ref{RS-F-Equations-Approximate}) can be immediately converted
into a $3 \times 3$ matrix representation.  However, this
representation does not contain the divergence conditions (the first
and the fourth Maxwell equations) contained in the second equation
in~(\ref{RS-F-Equations-Approximate}).  A further compactification is
possible only by expressing the Maxwell equations in a $4 \times 4$
matrix representation.  To this end,  using the components of the
Riemann-Silberstein vector, we define,
\bea
\Psi^{+} (\r , t)
& = &
\left[
\ba{c}
- F_x^{+}  + \i F_y^{+} \\
F_z^{+} \\
F_z^{+} \\
F_x^{+} + \i F_y^{+}
\ea
\right]\,, \quad \quad
\Psi^{-} (\r , t)
=
\left[
\ba{c}
- F_x^{-}  - \i F_y^{-} \\
F_z^{-} \\
F_z^{-} \\
F_x^{-} - \i F_y^{-}
\ea
\right]\,.
\eea
The vectors for the sources are
\bea
W^{+}
& = &
\left(\frac{1}{\sqrt{2 \epsilon}}\right)
\left[
\ba{c}
- J_x + \i J_y \\
J_z - v \rho \\
J_z + v \rho \\
J_x + \i J_y
\ea
\right]\,, \quad \quad
W^{-}
=
\left(\frac{1}{\sqrt{2 \epsilon}}\right)
\left[
\ba{c}
- J_x - \i J_y \\
J_z - v \rho \\
J_z + v \rho \\
J_x - \i J_y
\ea
\right]\,.
\label{Potentials-s}
\eea
Then we obtain
\bea
\frac{\partial}{\partial t}
\Psi^{+}
& = &
- v
\left\{ {\mbox{\boldmath $M$}} \cdot \Nab \right\} \Psi^{+}
- W^{+}\, \nn \\
\frac{\partial}{\partial t}
\Psi^{-}
& = &
- v
\left\{ {\mbox{\boldmath $M$}}^{*} \cdot \Nab \right\} \Psi^{-}
- W^{-}\,,
\label{Spinorial}
\eea
where `$^{*}$' denotes complex-conjugation and
the triplet,
$
{\mbox{\boldmath $M$}} = \left(M_x\,, M_y\,, M_z \right)
$
is expressed in terms of
\bea
\Omega
& = &
\left[
\ba{cc}
{\mbox{\boldmath $0$}} & - \one \\
\one & {\mbox{\boldmath $0$}}
\ea
\right]\,, \qquad
\beta =
\left[
\ba{cccc}
\one & {\mbox{\boldmath $0$}} \\
{\mbox{\boldmath $0$}} & - \one
\ea
\right]\,, \qquad
\one =
\left[
\ba{cc}
1 & 0 \\
0 & 1
\ea
\right]\,.
\label{}
\eea
Alternately, we may use the matrix $J = - \Omega$.  Both differ by
a sign.  For our purpose it is fine to use either $\Omega$ or $J$.
However, they have a different meaning: $J$ is {\em contravariant}
and $\Omega$ is {\em covariant}; The matrix $\Omega$ corresponds
to the Lagrange brackets of classical mechanics and $J$ corresponds
to the Poisson brackets.  An important relation is $\Omega = J^{-1}$.
The $M$-matrices are:
\bea
M_x & = &
\left[
\ba{cccc}
0 & 0 & 1 & 0 \\
0 & 0 & 0 & 1 \\
1 & 0 & 0 & 0 \\
0 & 1 & 0 & 0
\ea
\right] = - \beta \Omega  \,, \nn \\
M_y & = &
\left[
\ba{cccc}
0 & 0 & - \i & 0 \\
0 & 0 & 0 & - \i \\
\i & 0 & 0 & 0 \\
0 & \i & 0 & 0
\ea
\right] = \i \Omega \,, \nn \\
M_z & = &
\left[
\ba{cccc}
1 & 0 & 0 & 0 \\
0 & 1 & 0 & 0 \\
0 & 0 & -1 & 0 \\
0 & 0 & 0 & - 1
\ea
\right] = \beta \,.
\label{M-Matrices}
\eea
Each of the four Maxwell equations are easily obtained from the
matrix representation in~(\ref{Spinorial}).  This is done by taking
the sums and differences of row-I with row-IV and row-II with row-III
respectively.  The first three give the $y$, $x$ and $z$ components
of the curl and the last one gives the divergence conditions present
in the evolution equation~(\ref{RS-F-Equations-Approximate}).

It is to be noted that the matrices~${\mbox{\boldmath $M$}}$ are all
non-singular and all are hermitian.  Moreover, they satisfy the usual
algebra of the Dirac matrices, including,
\bea
M_x \beta = - \beta M_x\,, \nn \\
M_y \beta = - \beta M_y\,, \nn \\
M_x^2 = M_y^2 = M_z^2 = I\,, \nn \\
M_x M_y = - M_y M_x = \i M_z\,, \nn \\
M_y M_z = - M_z M_y = \i M_x\,, \nn \\
M_z M_x = - M_x M_z = \i M_y\,.
\label{Identities}
\eea

Before proceeding further we note the following: The pair
$(\Psi^{\pm} , {\mbox{\boldmath $M$}})$ are {\bf not} unique.
Different choices of $\Psi^{\pm}$ would give rise to different
${\mbox{\boldmath $M$}}$, such that the triplet
${\mbox{\boldmath $M$}}$ continues to to satisfy the algebra of the
Dirac matrices in~(\ref{Identities}).
We have preferred $\Psi^{\pm}$ {\em via} the the Riemann-Silberstein
vector~(\ref{R-S-Conjugate}) in~\cite{Birula}.  This vector has certain
advantages over the other possible choices.   The Riemann-Silberstein
vector is well-known in classical electrodynamics and has certain
interesting properties and uses~\cite{Birula}.

In deriving the above $4 \times 4$ matrix representation of the
Maxwell equations we have ignored the spatial and temporal derivatives
of $\epsilon (\r , t)$ and $\mu (\r , t)$ in the first two of the
Maxwell equations.  We have treated
$\epsilon$ and $\mu$ as `local' constants.

\section{Inhomogeneous Medium}
In the previous section we wrote the evolution equations for the
Riemann-Silberstein vector in~(\ref{RS-F-Equations-Approximate}), for
a medium, treating $\epsilon (\r , t)$ and $\mu (\r , t)$ as `local'
constants.  From these pairs of equations we wrote the matrix form of
the Maxwell equations.  In this section we shall write the exact
equations taking into account the spatial and temporal variations
of $\epsilon (\r , t)$ and $\mu (\r , t)$.  It is very much possible to
write the required evolution equations using $\epsilon (\r , t)$ and
$\mu (\r , t)$.  But we shall follow the procedure in~\cite{Birula} of
using the two derived {\em laboratory functions}
\bea
{\rm Velocity ~ Function:} \, v (\r , t)
& = &
\frac{1}{\sqrt{\epsilon (\r , t) \mu (\r , t)}} \nn \\
{\rm Resistance ~ Function:} \, h (\r , t)
& = &
\sqrt{\frac{\mu (\r , t)}{\epsilon (\r , t)}}\,.
\eea
The function, $v (\r , t)$ has the dimensions of velocity and the
function, $h (\r , t)$ has the dimensions of resistance (measured
in Ohms).  We can equivalently use the {\em Conductance Function},
$\kappa (\r , t) = {1}/{h (\r , t)} =
{\epsilon (\r , t)}/{\mu (\r , t)}$ (measured in Ohms$^{- 1}$
or Mhos!) in place of the resistance function, $h (\r , t)$.  These
derived functions enable us to understand the dependence of the
variations more transparently~\cite{Birula}.  Moreover the derived
functions are the ones which are measured experimentally.  In terms
of these functions,  $\epsilon = {1}/{\sqrt{v h}}$ and
$\mu = \sqrt{{h}/{v}}$.
Using these functions the exact equations satisfied by
$\F^{\pm} \left(\r , t \right)$ are
\bea
\i \frac{\partial }{\partial t} \F^{+} \left(\r , t \right)
& = &
v (\r , t) \left( \Nab \times \F^{+} \left(\r , t \right) \right)
+ \frac{1}{2}
\left(\Nab v (\r , t) \times \F^{+} \left(\r , t \right) \right) \nn \\
& &
+ \frac{v (\r , t)}{2 h (\r)}
\left(\Nab h (\r , t) \times \F^{-} \left(\r , t \right) \right)
- \frac{\i}{\sqrt{2}} \sqrt{v (\r , t) h (\r , t)}\,
{\mbox{\boldmath $J$}} \nn \\
& &
+ \frac{\i}{2} \frac{\dot{v} (\r , t)}{v (\r , t)}
\F^{+} \left(\r , t \right)
+ \frac{\i}{2} \frac{\dot{h} (\r , t)}{h (\r , t)}
\F^{-} \left(\r , t \right) \nn \\
\i \frac{\partial }{\partial t} \F^{-} \left(\r , t \right)
& = &
- v (\r , t) \left( \Nab \times \F^{-} \left(\r , t \right) \right)
- \frac{1}{2}
\left(\Nab v (\r , t) \times \F^{-} \left(\r , t \right) \right) \nn \\
& &
- \frac{v (\r , t)}{2 h (\r , t)}
\left(\Nab h (\r , t) \times \F^{+} \left(\r , t \right) \right)
- \frac{\i}{\sqrt{2}} \sqrt{v (\r , t) h (\r , t)}\,
{\mbox{\boldmath $J$}} \nn \\
& &
+ \frac{\i}{2} \frac{\dot{v} (\r , t)}{v (\r , t)}
\F^{-} \left(\r , t \right)
+ \frac{\i}{2} \frac{\dot{h} (\r , t)}{h (\r , t)}
\F^{+} \left(\r , t \right) \nn \\
\Nab \cdot \F^{+} \left(\r , t \right)
& = &
\frac{1}{2 v (\r , t)}
\left(\Nab v (\r , t) \cdot \F^{+} \left(\r , t \right) \right) \nn \\
& &
+
\frac{1}{2 h (\r , t)}
\left(\Nab h (\r , t) \cdot \F^{-} \left(\r , t \right) \right) \nn \\
& &
+ \frac{1}{\sqrt{2}} \sqrt{v (\r , t) h (\r , t)}\, \rho \,, \nn \\
\Nab \cdot \F^{-} \left(\r , t \right)
& = &
\frac{1}{2 v (\r , t)}
\left(\Nab v (\r , t) \cdot \F^{-} \left(\r , t \right) \right) \nn \\
& &
+
\frac{1}{2 h (\r , t)}
\left(\Nab h (\r , t) \cdot \F^{+} \left(\r , t \right) \right) \nn \\
& &
+ \frac{1}{\sqrt{2}} \sqrt{v (\r , t) h (\r , t)}\, \rho \,,
\label{RS-F-Equations-Exact}
\eea
where
$\dot{v} = \frac{\partial v}{\partial t}$ and
$\dot{h} = \frac{\partial h}{\partial t}$.
The evolution equations in~(\ref{RS-F-Equations-Exact}) are exact
(for a linear media) and the dependence on the variations
of $\epsilon (\r , t)$ and $\mu (\r , t)$ has been neatly expressed
through the two derived functions.   The  {\em coupling} between
$\F^{+} \left(\r , t \right)$ and $\F^{-} \left(\r , t \right)$.
is {\em via} the gradient and time-derivative of only one derived
function namely, $h (\r , t)$ or equivalently $\kappa (\r , t)$.
Either of these can be used and both are the directly measured
quantities.  We further note that the dependence of the coupling
is logarithmic
\bea
\frac{1}{h (\r , t)} \Nab h (\r , t)
=
\Nab \left\{\ln {\left(h (\r , t) \right)} \right\}\,, \quad
\frac{1}{h (\r , t)} \dot{h} (\r , t)
=
\frac{\partial }{\partial t}
\left\{\ln {\left(h (\r , t) \right)} \right\}\,,
\label{logarithm}
\eea
where `$\ln$' is the natural logarithm.

The coupling can be best summarized by expressing the equations
in~(\ref{RS-F-Equations-Exact}) in a (block) matrix form.
For this we introduce the following logarithmic function
\bea
{\cal L} (\r , t)
& = &
\frac{1}{2}
\left\{
\one \ln {\left(v (\r , t)\right)}
+
\sigma_x
\ln {\left(h (\r , t)\right)}
\right\}\,,
\label{logarithmic-function}
\eea
where
$\sigma_x$ is one the triplet of the Pauli matrices
\bea
{\mbox{\boldmath $\sigma$}}
& = &
\left[
\sigma_x =
\left[
\ba{cc}
0 & 1 \\
1 & 0
\ea
\right]\,, \
\sigma_y =
\left[
\ba{lr}
0 & - \i \\
\i & 0
\ea
\right]\,, \
\sigma_z =
\left[
\ba{lr}
1 & 0 \\
0 & -1
\ea
\right]
\right]\,.
\eea
Using the above notation the matrix form of the
equations in~(\ref{RS-F-Equations-Exact}) is
\bea
\i
\left\{
\one \frac{\partial }{\partial t}
- \frac{\partial }{\partial t} {\cal L}
\right\}
\left[
\ba{cc}
\F^{+} \left(\r , t \right) \\
\F^{-} \left(\r , t \right)
\ea
\right]
& = &
v (\r) \sigma_z
\left\{
\one \Nab + \Nab {\cal L}
\right\} \times
\left[
\ba{cc}
\F^{+} \left(\r , t \right) \\
\F^{-} \left(\r , t \right)
\ea
\right]\, \nn \\
& &
- \frac{\i}{\sqrt{2}} \sqrt{v (\r , t) h (\r , t)}\,
{\mbox{\boldmath $J$}} \nn \\
\left\{
\one \Nab - \Nab {\cal L}
\right\} \cdot
\left[
\ba{cc}
\F^{+} \left(\r , t \right) \\
\F^{-} \left(\r , t \right)
\ea
\right]
& = & + \frac{1}{\sqrt{2}} \sqrt{v (\r , t) h (\r , t)}\, \rho \,,
\label{six-evolution}
\eea
where the dot-product and the cross-product are to be
understood as
\bea
\left[
\ba{cc}
{\mbox{\boldmath $A$}} & {\mbox{\boldmath $B$}} \\
{\mbox{\boldmath $C$}} & {\mbox{\boldmath $D$}}
\ea
\right]
\cdot
\left[
\ba{c}
{\mbox{\boldmath $u$}} \\
{\mbox{\boldmath $v$}}
\ea
\right]
& = &
\left[
\ba{c}
{\mbox{\boldmath $A$}} \cdot {\mbox{\boldmath $u$}}
+
{\mbox{\boldmath $B$}} \cdot {\mbox{\boldmath $v$}} \\
{\mbox{\boldmath $C$}} \cdot {\mbox{\boldmath $u$}}
+
{\mbox{\boldmath $D$}} \cdot {\mbox{\boldmath $v$}}
\ea
\right] \nn \\
\left[
\ba{cc}
{\mbox{\boldmath $A$}} & {\mbox{\boldmath $B$}} \\
{\mbox{\boldmath $C$}} & {\mbox{\boldmath $D$}}
\ea
\right]
\times
\left[
\ba{c}
{\mbox{\boldmath $u$}} \\
{\mbox{\boldmath $v$}}
\ea
\right]
& = &
\left[
\ba{c}
{\mbox{\boldmath $A$}} \times {\mbox{\boldmath $u$}}
+
{\mbox{\boldmath $B$}} \times {\mbox{\boldmath $v$}} \\
{\mbox{\boldmath $C$}} \times {\mbox{\boldmath $u$}}
+
{\mbox{\boldmath $D$}} \times {\mbox{\boldmath $v$}}
\ea
\right]\,.
\eea
It is to be noted that the $6 \times 6$ matrices in the evolution
equations in~(\ref{six-evolution}) are either hermitian or
antihermitian.  Any dependence on the variations of
$\epsilon (\r , t)$ and $\mu (\r , t)$ is at best `weak'.  We further
note, $\Nab \left(\ln {\left(v (\r , t) \right)} \right)
= - \Nab \left(\ln {\left(n (\r , t) \right)} \right)$ and
$\frac{\partial }{\partial t}
\left(\ln {\left(v (\r , t) \right)} \right)
= - \frac{\partial }{\partial t}
\left(\ln {\left(n (\r , t) \right)} \right)$.
In some media, the coupling may vanish
($\Nab h (\r , t) = 0$ and $\dot{h} (\r , t) = 0$) and in the same
medium the refractive index, $n (\r , t) = {c}/{v (\r , t)}$ may
vary ($\Nab n (\r , t) \ne 0$ or/and $\dot{n} (\r , t) \ne 0$).
It may be further possible to use the approximations
$\Nab \left(\ln {\left(h (\r , t) \right)} \right)
\approx 0$ and
$\frac{\partial }{\partial t}
\left(\ln {\left(h (\r , t) \right)} \right)
\approx 0$.

We shall be using the following matrices to express the exact
representation
\bea
{\mbox{\boldmath $\Sigma$}}
=
\left[
\ba{cc}
{\mbox{\boldmath $\sigma$}} & {\mbox{\boldmath $0$}} \\
{\mbox{\boldmath $0$}} & {\mbox{\boldmath $\sigma$}}
\ea
\right]\,, \qquad
{\mbox{\boldmath $\alpha$}}
=
\left[
\ba{cc}
{\mbox{\boldmath $0$}} & {\mbox{\boldmath $\sigma$}} \\
{\mbox{\boldmath $\sigma$}} & {\mbox{\boldmath $0$}}
\ea
\right]\,, \qquad
{\mbox{\boldmath $I$}}
=
\left[
\ba{cc}
\one & {\mbox{\boldmath $0$}} \\
{\mbox{\boldmath $0$}} & \one
\ea
\right]\,,
\eea
where ${\mbox{\boldmath $\Sigma$}}$ are the Dirac spin matrices and
${\mbox{\boldmath $\alpha$}}$ are the matrices used in the Dirac
equation.  Then,
\bea
& &
\frac{\partial }{\partial t}
\left[
\ba{cc}
{\mbox{\boldmath $I$}} & {\mbox{\boldmath $0$}} \\
{\mbox{\boldmath $0$}} & {\mbox{\boldmath $I$}}
\ea
\right]
\left[
\ba{cc}
\Psi^{+} \\
\Psi^{-}
\ea
\right]
-
\frac{\dot{v} (\r , t)}{2 v (\r , t)}
\left[
\ba{cc}
{\mbox{\boldmath $I$}} & {\mbox{\boldmath $0$}} \\
{\mbox{\boldmath $0$}} & {\mbox{\boldmath $I$}}
\ea
\right]
\left[
\ba{cc}
\Psi^{+} \\
\Psi^{-}
\ea
\right] \nn \\
& & \qquad \qquad \quad
+
\frac{\dot{h} (\r , t)}{2 h (\r , t)}
\left[
\ba{cc}
{\mbox{\boldmath $0$}} & \i \beta \alpha_y \\
\i \beta \alpha_y & {\mbox{\boldmath $0$}}
\ea
\right]
\left[
\ba{cc}
\Psi^{+} \\
\Psi^{-}
\ea
\right]
\nn \\
& & \qquad \quad
=  - v (\r , t)
\left[
\ba{ccc}
\left\{
{\mbox{\boldmath $M$}} \cdot \Nab
+
{\mbox{\boldmath $\Sigma$}} \cdot {\mbox{\boldmath $u$}}
\right\}
& &
- \i \beta
\left({\mbox{\boldmath $\Sigma$}} \cdot {\mbox{\boldmath $w$}}\right)
\alpha_y
\\
- \i \beta
\left({\mbox{\boldmath $\Sigma$}}^{*} \cdot {\mbox{\boldmath $w$}}\right)
\alpha_y
& &
\left\{
{\mbox{\boldmath $M$}}^{*} \cdot \Nab
+
{\mbox{\boldmath $\Sigma$}}^{*} \cdot {\mbox{\boldmath $u$}}
\right\}
\ea
\right]
\left[
\ba{cc}
\Psi^{+} \\
\Psi^{-}
\ea
\right] \nn \\
& & \qquad \quad \quad
- \left[
\ba{cc}
{\mbox{\boldmath $I$}} & {\mbox{\boldmath $0$}} \\
{\mbox{\boldmath $0$}} & {\mbox{\boldmath $I$}}
\ea
\right]
\left[
\ba{c}
W^{+} \\
W^{-}
\ea
\right] 
\eea
where
\bea
{\mbox{\boldmath $u$}} (\r , t)
& = &
\frac{1}{2 v (\r , t)} \Nab v (\r , t)
=
\frac{1}{2} \Nab \left\{\ln v (\r , t) \right\}
=
- \frac{1}{2} \Nab \left\{\ln n (\r , t) \right\} \nn \\
{\mbox{\boldmath $w$}} (\r , t)
& = &
\frac{1}{2 h (\r , t)} \Nab h (\r , t)
=
\frac{1}{2} \Nab \left\{\ln h (\r , t) \right\}
\eea
The above representation contains thirteen $8 \times 8$ matrices!
Ten of these are hermitian.  The exceptional ones are the ones that
contain the three components of ${\mbox{\boldmath $w$}} (\r , t)$,
the logarithmic gradient of the resistance function.  These three
matrices are antihermitian.

\section{Concluding Remarks}
We have been able to express the Maxwell equations in a matrix form in
a medium with varying permittivity $\epsilon (\r , t)$ and permeability
$\mu (\r , t)$, in presence of sources.  We have been able to do so
using a single equation instead of a {\em pair} of matrix equations.
We have used $8 \times 8$ matrices and have been able to separate the
dependence of the coupling between the upper components ($\Psi^{+}$)
and the lower components ($\Psi^{-}$) through the two laboratory
functions.  Moreover the exact matrix representation has an algebraic
structure very similar to the Dirac equation.  We feel that this
representation would be more suitable for some of the studies related
to the {\em photon wave function}.  This representation is the starting
point for the
{\em exact formalism of Maxwell optics}~\cite{Khan-3}-\cite{Khan-5}.
This formalism provides a unified treatment of beam-optics and
polarization.

\end{document}